\documentclass{elsarticle}
\usepackage{amsmath,amssymb}
\usepackage{lscape}
\usepackage{hyperref}
\usepackage{multirow}

\newcommand{\be}{\begin{equation}}
\newcommand{\ee}{\end{equation}}
\newcommand{\bea}{\begin{eqnarray}}
\newcommand{\eea}{\end{eqnarray}}
\newcommand{\rbox}{\rule[-0.25cm]{0cm}{8mm}}

\begin{document}

\title{ Two photon couplings of the lightest isoscalars from BELLE data }

\author{ Ling-Yun Dai }
\ead{lingyun@jlab.org}
\author{ M.R. Pennington }
\ead{michaelp@jlab.org}

\address{Theory Center, Thomas Jefferson National Accelerator Facility,\\ Newport News, VA 23606, USA}

\begin{abstract}
Amplitude Analysis of two photon production of $\pi\pi$ and ${\overline K}K$, using
S-matrix constraints and fitting all available data, including the latest precision results from Belle,
yields a single partial wave solution up to 1.4~GeV. The two photon couplings of the $\sigma/f_0(500)$, $f_0(980)$ and $f_2(1270)$
are determined from the residues of the resonance poles.
These amplitudes are a key input into the newly developed  dispersive approach to calculating hadronic light-by-light scattering for $(g-2)$ of the muon.
\end{abstract}

\maketitle

\section{Introduction}
Two photon reactions play a special role in the study of QCD: photons pick out the charged components of hadrons and so probe their structure. In this Letter we present the results of a comprehensive Amplitude Analysis of all data on $\gamma\gamma\to\pi^+\pi^-$, $\pi^0\pi^0$, ${\overline K}K$ up to 1.4 GeV. This includes for the first time the high statistics data from  Belle on $\pi^+\pi^-$~\cite{Belle-pm}, $\pi^0\pi^0$~\cite{Belle-nn} and the very new $K_sK_s$~\cite{Belle-KsKs} channels in a coupled channel analysis.
The data have limited angular coverage and no polarization information. Nevertheless, unitarity links these two photon reactions to the corresponding meson-meson scattering processes. When combined with the other basic S-matrix principles of analyticity and crossing, these constraints make up for the limitations of the data, and make an Amplitude Analysis feasible. At present this can be implemented where the $\pi\pi$ and ${\overline K}K$ saturate unitarity, which is roughly up to 1.4-1.5 GeV. At higher energies multi-meson production becomes important, for which we do not yet have precise enough information to extend the analysis further.

Unitarity provides the main constraint on the determination of the partial wave amplitudes. For each amplitude with definite spin $J$, helicity $\lambda$ and isospin $I$, unitarity for the two photon process to hadrons requires
\bea\label{eq:Funi}
{\rm Im}\,F^I_{J\lambda}(\gamma\gamma\to \pi\pi;s)&=& \sum_i\,\rho_i(s)\,{F^I_{J\lambda}}^*(\gamma\gamma\to i;s)\, \cdot\, T^I_J(i \to \pi\pi; s) \quad ,
\eea
where $s$ is the square of the c.m. energy, $\rho_i$ is the standard phase space for channel $i$ and the sum is over all open channels. The hadronic amplitudes for reaction $i \to j$, $T^I_J(i\to j; s)$, themselves satisfy  partial wave unitarity, so that
\bea\label{eq:Tuni}
{\rm Im}\,T^I_{J}(i\to\pi\pi;s)&=& \sum_i\,\rho_i(s)\,{T^I_J}^*(i\to k;s)\, \cdot\, T^I_J(k \to \pi\pi; s)\quad ,
\eea
with a slightly more complicated form if the final state particles have spin. 
The relations set out in Eqs.~(\ref{eq:Funi},\ref{eq:Tuni}) are fulfilled by the simple condition:
\bea\label{eq:F}
F^I_{J\lambda}(\gamma\gamma\to\pi\pi)&=& \sum_i\; {\alpha_i}^I_{J\lambda}(s)\cdot T^I_J(i\to\pi\pi; s)\quad,
\eea
with obvious generalizations from $\pi\pi$, to ${\overline K}K$ we need here, and to any other final states. To satisfy the unitarity relations, Eqs.~(\ref{eq:Funi},\ref{eq:Tuni}),
the coupling functions $\alpha_i^{I,J,\lambda}(s)$ in Eq.~(\ref{eq:F}) must be real for real values of the c.m. energy $\sqrt{s}$ above the lowest threshold. Importantly, these functions only have left hand cuts, thereby ensuring that the two photon amplitudes have the same right hand cut structure as the hadronic amplitudes, as required by unitarity. This separation of $\alpha$ and $T$-matrix elements is  closely related to the $N$-$D$ separation of the $N/D$ method. However, it  is far simpler in practice to impose Eq.~(\ref{eq:F}) in an analysis of experimental data. Once zeros of  the hadron scattering amplitudes are divided out, the coupling functions, $\alpha(s)$, are readily parametrized by polynomials over the limited energy region we consider here. For a larger energy domain a conformal mapping would be more efficient. The zeros that are divided out are: for the $S$-waves the process-dependent Adler zeros of pseudoscalar scattering, and for higher waves the usual angular momentum threshold factors, reflecting the difference in threshold behavior between the hadron reactions and the interaction of spin-1 photons\footnote{In general one has to divide out any zeros of the sub-determinants of the $T$-matrix\cite{AMP-FSI}. }.

Importantly, the unitarity constraint given by Eq.~(\ref{eq:F}) ensures that Watson's theorem is fulfilled in the region of elastic unitarity, when all amplitudes with $\pi\pi$ final states in the same quantum numbers have the same phase. It also ensures that the poles of the hadronic $T$-matrix transmit to the two photon reaction in exactly the same positions.
For the constraint implied by Eq.~(\ref{eq:F}) to be used, one needs, of course, detailed information on the hadronic $T$-matrix elements. Considerable progress has been made over the past decades in refining knowledge of the key meson-meson scattering amplitudes for $\pi\pi\to\pi\pi$ and $\pi\pi\to {\overline K}K$. This has come about by new experimental information on near threshold $\pi\pi$ scattering from $K\to (\pi\pi)e\nu$ from NA48-2~\cite{NA48} (with input too from the DIRAC experiment~\cite{DIRAC}) at CERN, combined with data from the classic meson-meson scattering experiments on $\pi\pi\to\pi\pi$ from CERN-Munich~\cite{CERN-Munich}, and the $\pi\pi\to{\overline K}K$ from Argonne~\cite{Cohen80} and Brookhaven~\cite{Etkin82}. The hadronic amplitudes we need have been constructed by incorporating these data in dispersive analyses as recently done in~\cite{KPY,Descotes04}.  An outcome of this is a rather precise knowledge of the hadronic scattering amplitudes and in turn of pole positions of the key resonances in the energy region studied. These poles are automatically built into our two photon amplitudes, not through some simplistic Breit-Wigner forms, but through the detailed parametrization of the underlying meson-meson scattering amplitudes in Eq.~(\ref{eq:F}).
The positions of the dominant poles are listed in Table I. Their uncertainties are typically $\pm 10$~MeV or less in both the real and imaginary parts.

These scattering amplitudes embodied in a $K$-matrix framework, are  key inputs into unitarity for the two photon production of these same hadronic final states.  As already mentioned, the two photon data have limited angular coverage, only 60\% for the charged pions and 80\% for neutral, and no polarization information make an Amplitude Analysis challenging. As set out in Ref.~\cite{MRP8891}, the partial wave amplitudes are anchored at low energy by the fact that they can be accurately calculated close to threshold by the use not just of unitarity, Eqs.~(\ref{eq:Funi},\ref{eq:F}), but by the application of the dispersion relations to the two photon partial waves imposing the low energy theorem of Compton scattering.
The uncertainties in these calculations of the absolute two photon cross-section increase with increasing energy as shown in Refs.~\cite{MRP8891}-\cite{Phillips11}.
In a longer paper~\cite{DLY-MRP14}, we confirm that below 600 MeV the partial wave amplitudes are calculationally under good control. Above that energy we rely entirely on the experimental data constrained by the unitarity relation, Eq.~(\ref{eq:Funi}), implemented using Eq.~(\ref{eq:F}), to determine the possible amplitudes.

When this method was applied 25 years ago to the then available two photon data on $\pi\pi$ production  from SLAC and DESY~\cite{MarkII}-\cite{CB92}.
and with more uncertain hadronic inputs, several distinct classes of  two photon to $\pi\pi$ solutions were possible.
These had the $f_0(980)$ appearing as a peak of different sizes, or as a dip structure (as in $\pi\pi\to\pi\pi$) of different sizes, and with a range of helicity zero and two components for the $f_2(1270)$ (solutions A-E in~\cite{MRP90}, 1 and 2 in \cite{MRP98}). The new data from Belle on $\pi^+\pi^-$ in 5 MeV bins displayed the peaking of the $f_0(980)$, but still admitted a range of solutions, A and B in \cite{MRP08}).
With the two photon results from the $B$-factories, showing a clear structure for the $f_0(980)$, Figs.~1,2, we need to ensure our underlying meson-meson scattering amplitudes have this resonance built in correctly. While some aspects of the $f_0(980)$ are constrained by the dispersive analyses mentioned above~\cite{KPY,Descotes04}, these ignore the isospin breaking engendered by the kaon mass difference. With Belle providing two photon results on $\pi^+\pi^-$ production in 5~MeV bins, it is essential that our hadronic amplitudes also take into account the 8 MeV mass splitting between $K^+K^-$ and ${\bar K^0}K^0$ thresholds. This we do by requiring our hadronic amplitudes also fit the results of partial wave analyses of the BaBar results on $D_s$ decay into $S$-wave di-pion and di-kaon systems~\cite{BABAR-pi,BABAR-K}.

With the constraint of unitarity encoded in Eq.~(3) and the low partial waves anchored at low energy by dispersive constraints,
the present 3000 two photon data points for the $\pi\pi$ channels and 350 data on ${\overline K}K$, both integrated and differential cross-sections, are fitted. \begin{figure}[htbp]
\vspace{-1mm}
\includegraphics[width=1.0\textwidth,height=0.3\textheight]{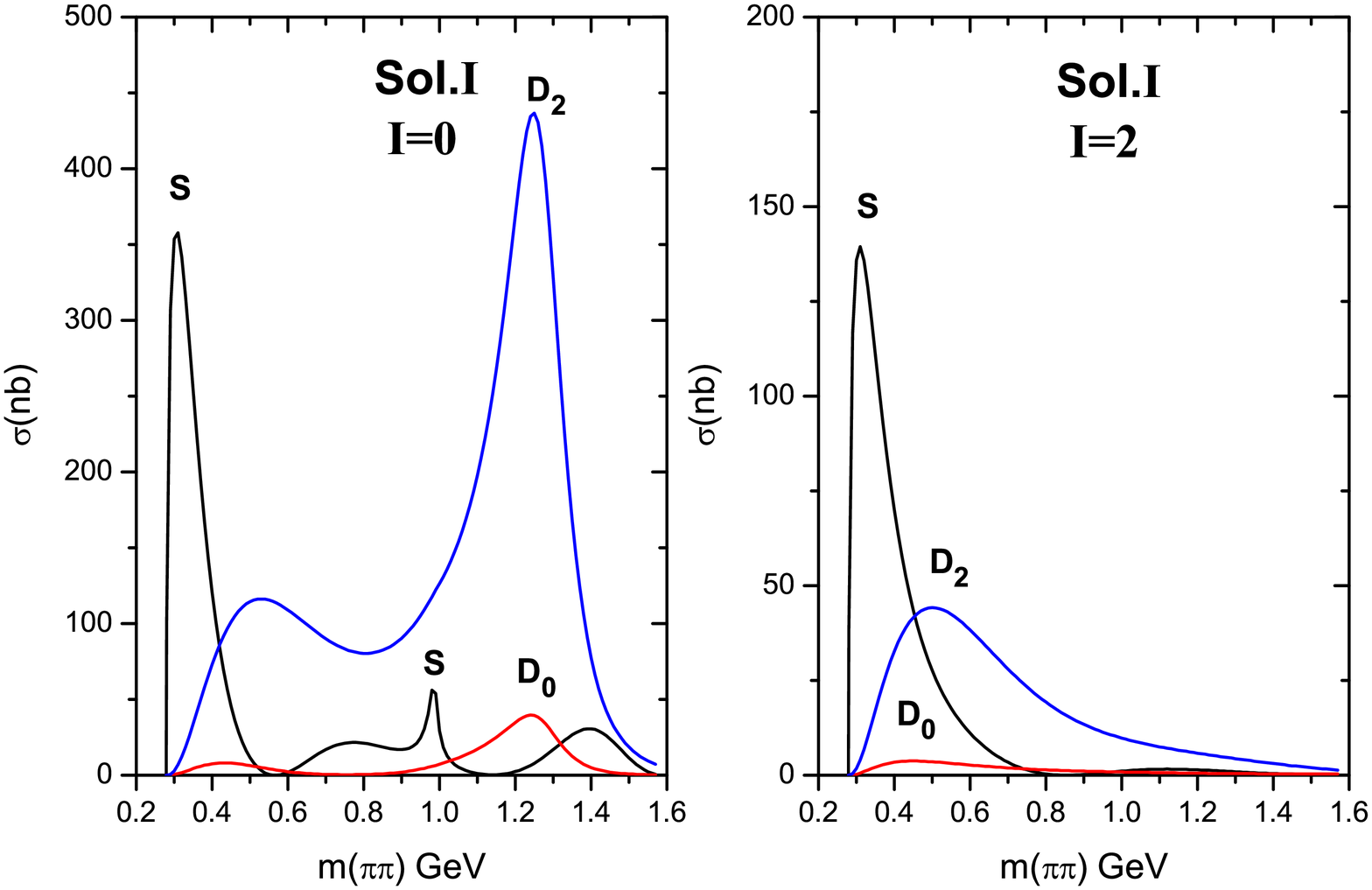}
\caption{\label{fig:csiso;pi} Individual partial wave components of the $\gamma\gamma\to\pi\pi$ integrated cross-section. }
\end{figure}
It is the addition of the $\pi^0\pi^0$ and ${\overline K}K$ results from Belle, particularly their latest $K_sK_s$ data, which are the first with accurate coverage from threshold upwards with angular information out to $\cos \theta =0.6-0.8$, that dramatically reduces the range of solutions in this coupled channel analysis below 1.5 GeV to the single solution (Solution I) presented here.  Above that energy the addition of multi-pion production information would be absolutely crucial. The partial wave decomposition of such reactions is unfortunately missing in the hadronic scattering sector.

Each data set has systematic uncertainties. In the case of that from Cello, these have been folded with the statistical errors  in their publication. In other cases, like that of Mark II and Belle $\pi\pi$ cross-sections, the main systematic uncertainty is in the absolute normalization of the cross-sections. A possible shift in normalization beyond 700~MeV between experimental datasets is included in our fitting procedure. The Belle charged pion results are used to set the scale. Their systematic shift of $\sim\pm 5\%$ should then be assigned to our solutions.

\section{Two photon couplings}
\begin{figure}[htbp]
\includegraphics[width=1.0\textwidth,height=0.25\textheight]{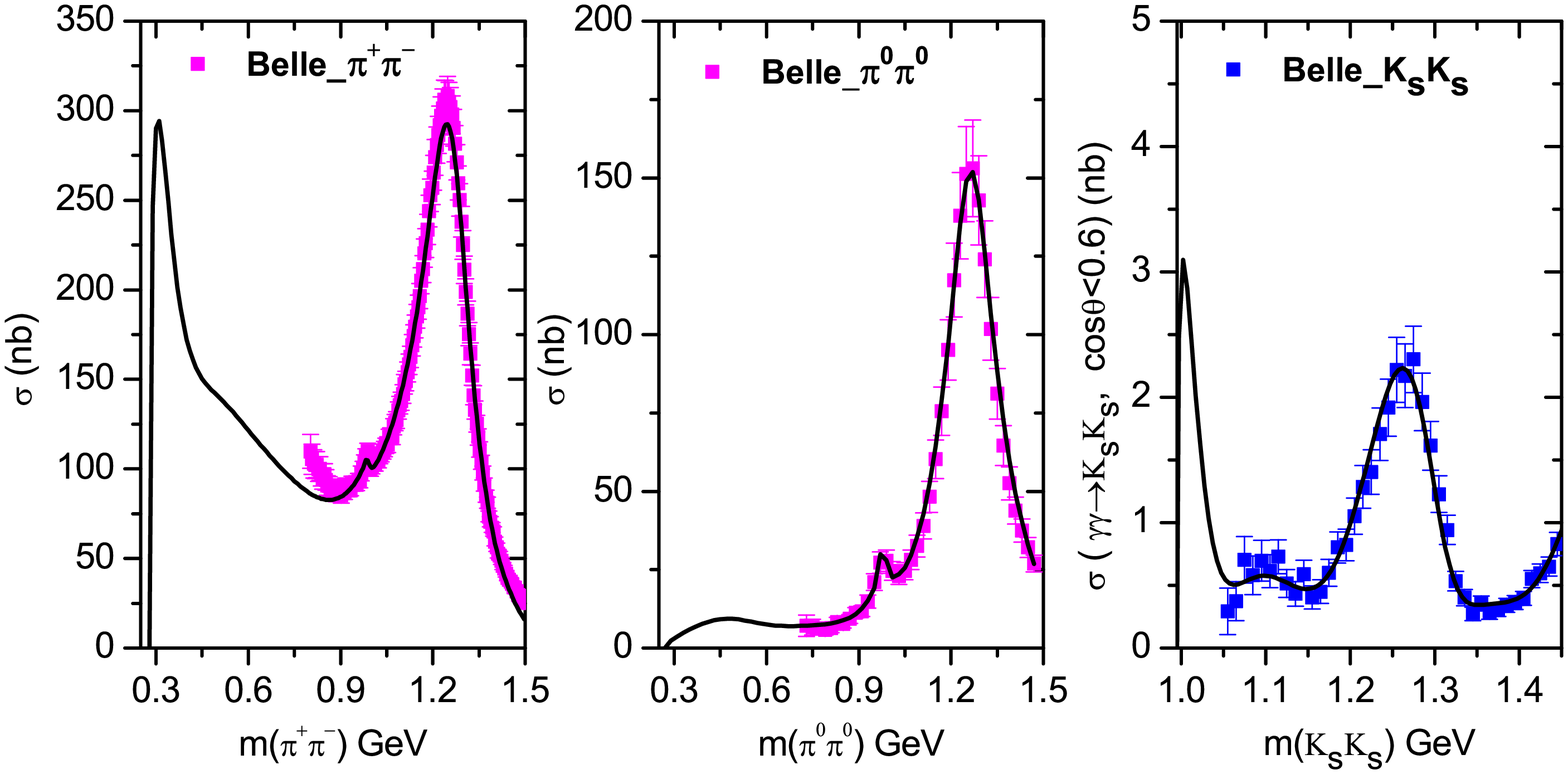}
\caption{\label{fig:cs;int} Solution.~I compared with the integrated cross-section datasets of Belle. The $\gamma\gamma\rightarrow\pi^+\pi^-$ process~\cite{Belle-pm} are integrated over $|\cos \theta|\,\le\,0.6$, $\gamma\gamma\rightarrow\pi^0\pi^0$~\cite{Belle-nn} is 0.7 and for $\gamma\gamma\rightarrow K_sK_s$~\cite{Belle-KsKs} it is 0.6. }
\end{figure}
Having data on both the charged and neutral pion final states allows a separation of the $I=0$ and 2  components of the $\gamma\gamma\to\pi\pi$ amplitudes,
and the determination of the individual partial waves with helicity-0 and 2 within narrower ranges than previously possible. The $\pi\pi$ partial wave cross-sections for $J=0,2$ are shown in Fig.~\ref{fig:csiso;pi}.
How these describe the Belle integrated cross-sections is shown in Fig.~\ref{fig:cs;int} for $\pi^+\pi^-$, $\pi^0\pi^0$ and $K_sK_s$ production.
While only the comparison with Belle data are shown here, our Amplitudes describe all the available data from Mark~II, CELLO, Crystal Ball, TASSO, ARGUS and TPC~\cite{MarkII}-\cite{CB92}, \cite{ARGUS89}-\cite{TASSO86} too.

The publication of the Belle $\pi^0\pi^0$ results have highlighted some systematic \lq\lq imperfections'' in the Belle $\pi^+\pi^-$ data, already apparent when compared with Mark II and Cello results. The charged particle mode is dominated by $\mu^+\mu^-$ production by orders of magnitude. Belle present results above 800~MeV where they believe they can separate $\mu$'s from $\pi$'s. This may not be correct with their acceptance, as their data have a strange angular dependence around $\cos \theta \sim 0.6$ below 1~GeV, which in turn produces the upward sweep of the integrated cross-section seen in Fig.~\ref{fig:cs;int}, not found by Mark II and Cello, discussed further in ~\cite{DLY-MRP14}. Which $\pi^+\pi^-$ results are correct will be checked by a forthcoming measurement by KLOE-II at DAPHNE~\cite{MRP-KLOE2,KLOE2}.

\begin{figure}[htbp]
\includegraphics[width=1.0\textwidth,height=0.4\textheight]{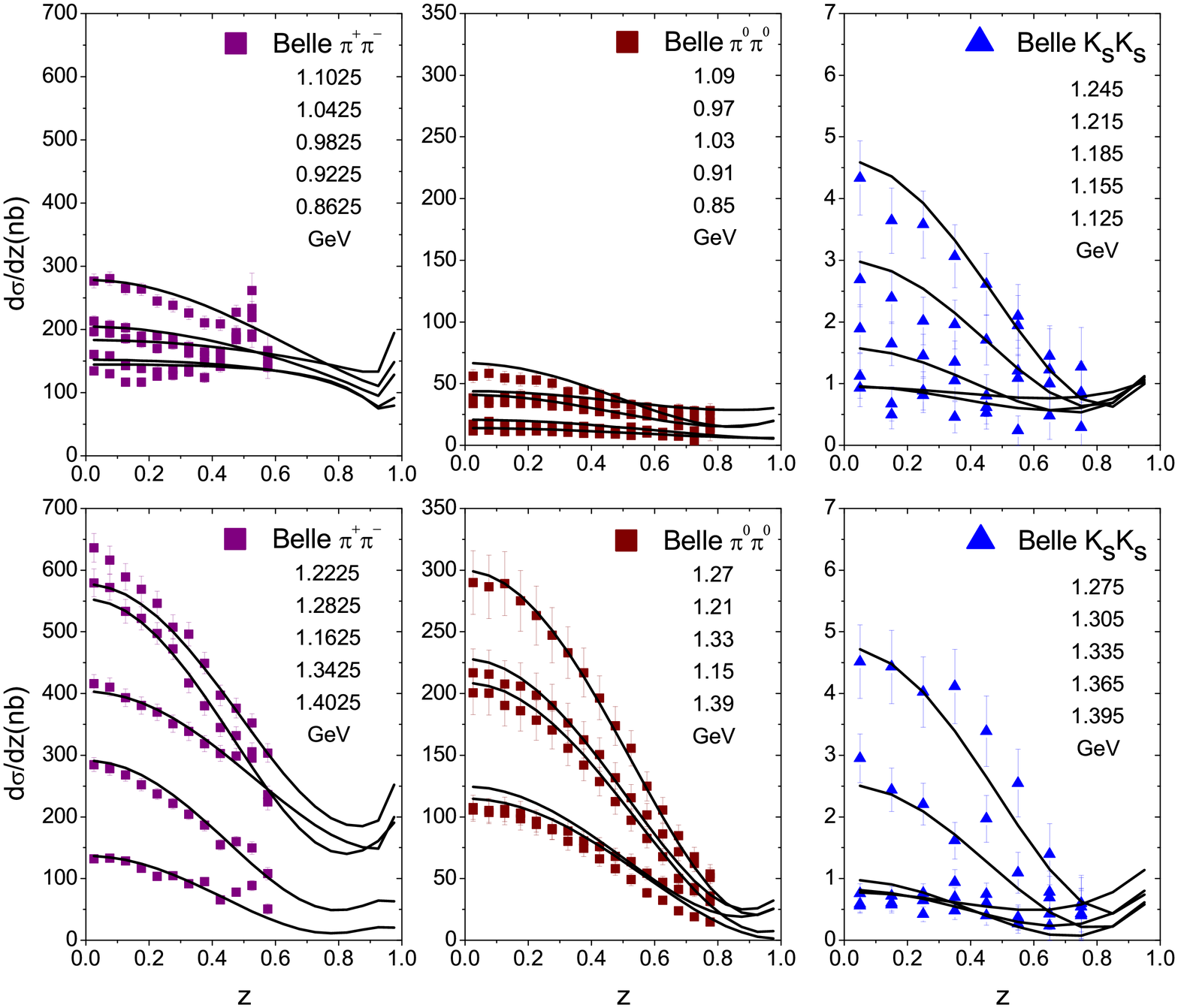}
\caption{\label{fig:dcs} Solution.~I compared with the differential cross-section datasets of Belle. The $\gamma\gamma\rightarrow\pi^+\pi^-$ process is from~\cite{Belle-pm}, $\gamma\gamma\rightarrow\pi^0\pi^0$ from~\cite{Belle-nn} and for $\gamma\gamma\rightarrow K_sK_s$ from~\cite{Belle-KsKs}. }
\end{figure}

While only even isospins occur for the $\pi\pi$ channel, the $K^+K^-$ and ${\overline K}^0K^0$ have $I=0,1$. The isoscalar channels are highly constrained by unitarity. However, the isovector channel has to be freely parametrized. Nevertheless, the input of the ${\overline K}K$ data fixes the isoscalar partial waves. The way our amplitudes describe the angular distributions for $\pi^+\pi^-$, $\pi^0\pi^0$ and $K_sK_s$ is illustrated in Fig.~\ref{fig:dcs} at a number of representative energies.
The complete datasets used, the treatment of systematic errors and the dispersive technology used, together with the full results are described fully in a longer paper~\cite{DLY-MRP14}.

The outcome of this analysis is the set of  partial wave amplitudes, the cross sections for which are shown in Fig.~\ref{fig:csiso;pi}. In turn, these fix the two photon couplings of the resonance poles that occur in these channels. These are dominated by the $\sigma/f_0(500)$, $f_0(980)$ and $f_2(1270)$.  As mentioned already these have been determined to $\pm 10$~MeV for the $\sigma$ and $f_2$, and to $\pm 3$~MeV for the $f_0(980)$ by the analyses of the hadronic scattering amplitudes. The residues of these poles  on the appropriate nearby sheet of the energy plane determine the two photon coupling $g_{\gamma\gamma}$ for each state. The two photon width, $\Gamma(R\to\gamma\gamma)$, is readily defined for an isolated, narrow state with a nearby pole in the complex energy plane, well-separated from threshold cuts. For the states that dominate the channels studied here, that are broad and overlapping each other with strongly coupled thresholds, we still use the same definition, viz:
\be\label{eq:photonwidth}
\Gamma(R\to\gamma\gamma)= \frac {\alpha^2}{4~(2J+1)~m_R}\,|g_{\gamma\gamma}|^2\;,
\ee
where $\alpha$ is the usual QED fine structure constant, $J$ is the spin of the resonance, and $m_R$ its mass. Here we take $m_R$ to be the modulus of the pole position in the energy plane.
Other definitions are folded into the uncertainties discussed below. This $\Gamma(R\to\gamma\gamma)$ is, of course, {\bf not} a physical quantity, but merely an intuitive way of re-expressing $|g_{\gamma\gamma}|$.
These values are listed in Table~1. The uncertainties of the residues are from the two photon amplitudes, see Eq.~(\ref{eq:F}), the error being mainly caused by the uncertainties in the ${\alpha_i}^I_{J\lambda}(s)$.
The $T$-matrix elements contribute to the errors too, but these add only a few percent from the pole locations and the $\pi\pi$ couplings.

\begin{table}[htbp]
\begin{center}
{\footnotesize
\begin{tabular}{|c|c||c|c|c|c|c|c|}
\hline
\rule[-0.5cm]{0cm}{1cm}\multirow{2}{*}{\rule[-1cm]{0cm}{2cm}State} & \multirow{2}{*}{\rule[-1cm]{0cm}{2cm}Sh} &  pole locations
                & \multicolumn{3}{c|}{$g_{\gamma\gamma}=|g|e^{i\varphi}$}  & $\Gamma(f_J\to\gamma\gamma)$        &  $\lambda=0$  \\
\cline{4-6}
\rule[-0.5cm]{0cm}{1cm} & \multirow{2}{*}{}  & (GeV)  & $J_\lambda$ & $|g|~(GeV)$ & $\varphi$~($^\circ$) & (keV) &  fraction~\% \\
\hline\hline
\multirow{2}{*}{\rbox $f_2(1270)$}   & \multirow{2}{*}{\rbox III}  &  \multirow{2}{*}{\rbox  $1.267 -i0.108$}
                & \rbox $D_0$  & 0.35$\pm$0.03    &  168$\pm$6     &  \multirow{2}{*}{ \rbox 2.93$\pm$0.40 }    & \multirow{2}{*}{\rbox 8.7$\pm$1.7}\\
\cline{4-6}
\multirow{2}{*}{}                  & \multirow{2}{*}{}           &  \multirow{2}{*}{}
                & \rbox $D_2$   & 1.13$\pm$0.08    &  173$\pm$6     & \multirow{2}{*}{}                          & \multirow{2}{*}{}  \\
\hline
$\sigma/f_0(500)$     & \rbox II   & $0.441 -i0.272$   & S    & 0.26$\pm$0.01  & 105$\pm$3  &   2.05$\pm$0.21  &     100     \\
\hline
$f_0(980)$     & \rbox II   & $0.998 - i0.021$  & S    & 0.16$\pm$0.01  & -175$\pm$5      &   0.32$\pm$0.05  &     100     \\
\hline
\end{tabular}
\caption{\label{tab:poles} The isoscalar resonance poles and their two photon residues (both magnitude and phase) from our Amplitude Solution  are listed. The pole positions for the $\sigma$ and $f_2(1270)$ have an uncertainty of $\pm 10$~MeV for the real and imaginary parts, while for the $f_0(980)$ the errors are $\pm 3$~MeV. The two photon residues can be interpreted in terms of  two-photon partial widths using Eq.~(1). These are tabulated in keV. For each the fraction of the width provided by helicity zero is given:
for the scalar resonances, it is, of course, 100\%.}
}
\end{center}
\end{table}

Let us emphasise that the present work is the only  robust determination of the two photon couplings of the $f_0(980)$ and $f_2(1270)$ from a partial wave analysis that genuinely separates the $S$-waves from the $D$-waves, and $D$-waves with helicity two from that with helicity zero. Indeed, the values given in Table I are specified from the residues of the resonance poles rather than resonance plus background fits to data on a single charged channel, as for instance published by Belle~\cite{Belle-pm}. Continuing to the pole is the only rigorous way to determine resonance parameters. This is particularly apparent for the $\sigma$ with its very deep pole. The PDG values for its two photon \lq\lq width'' follow from determinations that use the method we have advocated~\cite{MRP06}, implemented by others~\cite{Oller0708,MRP08,yumao09}, and updated here.
That results now converge is reassuring.

\section{Discussion}
Model calculations have been made for these states depending on their \lq\lq primordial'' composition. How these are related to those in the real world of important meson final state interactions do not yet exist beyond models.
Kaon loop modelling by Achasov and collaborators~\cite{Achasov07} favors a tetraquark composition for the $f_0(980)$ with a $\gamma\gamma$ width predicting $\sim 270$~eV~\cite{Achasov82}. This is not very different from the prediction for a largely $K\overline{K}$ composition for the $f_0(980)$ by Hanhart {\it et al.}~\cite{Hanhart07} of 220~eV. Both model predictions are reasonably close to our extracted result of $(320\pm 50)$~eV, but quite different from the older prediction of  Barnes~\cite{BarnesKK} of $\sim 600$~eV  in the molecular model of Weinstein and Isgur~\cite{Weinstein}. A genuine strong coupling QCD calculation would clearly help here. Incidently, our \lq\lq opinion'' favours the  ${\overline K}K$ molecular structure as more appropriate, see ~\cite{MRP05,MRP10,Wilson}. It is such considerations that make a comparison with the two photon production of the $a_0(980)$ of special interest. However, results of comparable precision for isovector states must await a corresponding coupled channel analysis combining data on $\gamma\gamma\to\pi^0\eta$, ${K^+K^-}$ and ${\overline K^0}K^0$ with that on $\pi\pi$. While the two photon production of $\pi\pi$ and $\eta\pi$ channels, of course, access different isospins, the ${\overline K}K$ channels involve both $I=0,1$. Thus a larger global analysis would be required, which would inevitably involve multi-pion channels too. This is beyond our present ambitions.

Other analyses have combined dispersion relations with unitarity and hadronic scattering information, with the same basic philosophy as we have followed here. Calculations by Garcia-Martin and Moussallam~\cite{Moussallam10} have assumed that the crossed-channel exchanges, namely states in $\gamma\pi$ scattering, have known couplings and hence the direct channel $\gamma\gamma\to\pi\pi$ cross-sections can be predicted up to at least 1 GeV. As we shall discuss in a separate, more technical paper, single particle exchange (beyond the crucial one pion exchange of the Born amplitude) is likely a poor approximation to the multi-meson exchanges that control the details of the left hand cut amplitude. Hoferichter, Phillips and Schat~\cite{Phillips11} have used Roy-Steiner equations, deduced from dispersion relations on hyperbolae, to constrain the $\gamma\gamma\to\pi\pi$ amplitudes. Their analysis does not attempt to fit experimental information beyond 1~GeV directly, and they assume for instance that the input of $f_2(1270)$ only has helicity two couplings. Here we perform an Amplitude Analysis within a corresponding S-matrix framework, but in which data are used directly to determine the partial waves. From these we then determine the $\gamma\gamma$ couplings of each resonant pole.

The recent development of a dispersive approach to calculating hadronic light-by-light scattering to $(g-2)$ of the muon requires as input knowledge of two photon production of hadrons, of which $\pi\pi$ and ${\overline K}K$ are likely the most important. The amplitudes presented here for on-shell photons are thus a key component of a  robust determination of these contributions, as well as their uncertainties, critical for interpreting the present BNL measurement~\cite{BNL-g2} and assessing the prospects for the future Fermilab experiment~\cite{Fermilab-g2}.

\section*{Acknowledgements}

MRP thanks Yasushi Watanabe and Sadaharu Uehara for early access to the Belle $\pi^0\pi^0$ data that got this analysis started.
This paper has been authored by Jefferson Science Associates, LLC under U.S. DOE Contract No. DE-AC05-06OR23177.

\clearpage

\end{document}